\pdfoutput=1
\documentclass[fleqn,usenatbib,usedcolumn]{mnras}
\usepackage[british]{babel}             
\let\la=\lesssim 
\usepackage[T1]{fontenc}                
\usepackage{graphicx}                   
\usepackage[usenames,dvipsnames]{color}

\hypersetup{pdfauthor={I. Heywood},
               pdftitle={Field sources near the southern-sky calibrator PKS B1934-638: effect on spectral line observations with SKA-MID and its precursors},
               pdfkeywords={techniques: interferometric},
               bookmarksnumbered=true}

\volume{{\rm in press}}   

\usepackage{graphicx}	
\usepackage{amsmath}	
\usepackage{amssymb}	
\usepackage{multicol}        
\usepackage{bm}		
\usepackage{pdflscape}	
\usepackage{color}

\usepackage[T1]{fontenc}
\usepackage{ae,aecompl}

\usepackage{txfonts}

\title[Field sources near the southern-sky calibrator PKS B1934-638]{Field sources near the southern-sky calibrator PKS B1934-638: effect on spectral line observations with SKA-MID and its precursors}

\author[Heywood et al.]
{\parbox{\textwidth}{
\begin{flushleft}
I. Heywood$^{1,2,3}$\thanks{E-mail: ian.heywood@physics.ox.ac.uk}, 
E. Lenc$^{4}$, 
P. Serra$^{5}$, 
B. Hugo$^{3,2}$, 
K.~W.~Bannister$^{4}$, 
M.~E.~Bell$^{6}$, 
A.~Chippendale$^{4}$, 
L.~Harvey-Smith$^{7,8}$, 
J.~Marvil$^{9}$, 
D. McConnell$^{4}$ 
and M.~A.~Voronkov$^{4}$\\
\end{flushleft}
}
\footnotesize
\\
$^{1}$Astrophysics, Department of Physics, University of Oxford, Keble Road, Oxford OX1 3RH, UK\\
$^{2}$Department of Physics and Electronics, Rhodes University, PO Box 94, Grahamstown, 6140, South Africa\\
$^{3}$South African Radio Astronomy Observatory, 2 Fir Street, Black River Park, Observatory, Cape Town 7925, South Africa\\
$^{4}$CSIRO Astronomy and Space Science, PO Box 76, Epping, NSW 1710, Australia\\
$^{5}$INAF - Osservatorio Astronomico di Cagliari, Via della Scienza 5, I-09047 Selargius (CA), Italy\\
$^{6}$School of Mathematical and Physical Sciences, University of Technology Sydney, 15 Broadway, Ultimo, NSW 2007, Australia\\
$^{7}$School of Physics, University of New South Wales, Sydney, NSW 2052, Australia\\
$^{8}$Western Sydney University, Locked Bag 1797, Penrith, NSW, 2751, Australia\\
$^{9}$National Radio Astronomy Observatory, PO Box 0, Soccoro, NM 87801, USA\\}

\date{Accepted 2020 April 01. Received 2020 April 01; in original form 2019 July 13}

\pubyear{2020}

\begin{document}
\label{firstpage}
\pagerange{\pageref{firstpage}--\pageref{lastpage}}
\maketitle

\begin{abstract}

Accurate instrumental bandpass corrections are essential for the reliable interpretation of spectral lines from targeted and survey-mode observations with radio interferometers. Bandpass correction is typically performed by comparing measurements of a strong calibrator source to an assumed model, typically an isolated point source. The wide field-of-view and high sensitivity of modern interferometers means that additional sources are often detected in observations of calibrators. This can introduce errors into bandpass corrections and subsequently the target data if not properly accounted for. Focusing on the standard calibrator PKS B1934-638, we perform simulations to asses this effect by constructing a wide-field sky model. The cases of ASKAP (0.7--1.9 GHz), MeerKAT (UHF: 0.58--1.05 GHz; L-band: 0.87--1.67 GHz) and Band 2 (0.95--1.76 GHz) of SKA-MID are examined. The use of a central point source model during bandpass calibration is found to impart amplitude errors into spectra measured by the precursor instruments at the $\sim$0.2--0.5\% level dropping to $\sim$0.01\% in the case of SKA-MID. This manifests itself as ripples in the source spectrum, the behaviour of which is coupled to the distribution of the array baselines, the solution interval, the primary beam size, the hour-angle of the calibration scan, as well as the weights used when imaging the target. Calibration pipelines should routinely employ complete field models for standard calibrators to remove this potentially destructive contaminant from the data, a recommendation we validate by comparing our simulation results to a MeerKAT scan of PKS B1934-638, calibrated with and without our expanded sky model.

\end{abstract}





\section{Introduction}

The radio source PKS B1934-638 (J2000 19h39m25.02671s -63d42m45.6255s) is a compact steep spectrum source \citep[see e.g.][]{odea98} associated with Seyfert 2 type galaxy at a redshift of $z$~=~0.183 \citep{holt08}. It is heavily relied upon as a primary calibrator source for radio interferometers in the southern hemisphere as it is extremely stable \citep{ojha04}, exhibits no structure on scales larger than $\sim$40 milli-arcseconds \citep{tzioumis10}, and is effectively unpolarized (Hugo, priv. comm.), with limits:
\begin{equation}
\nonumber
\frac{Q^{2} + U^{2}}{I^{2}} \approx -32 ~\mathrm{dB}
\end{equation}
\noindent
and
\begin{equation}
\nonumber
\frac{Q^{2} + U^{2} + V^{2}}{I^{2}} \approx -30~\mathrm{dB}.
\end{equation}

\begin{figure}
\centering
\includegraphics[width=\columnwidth]{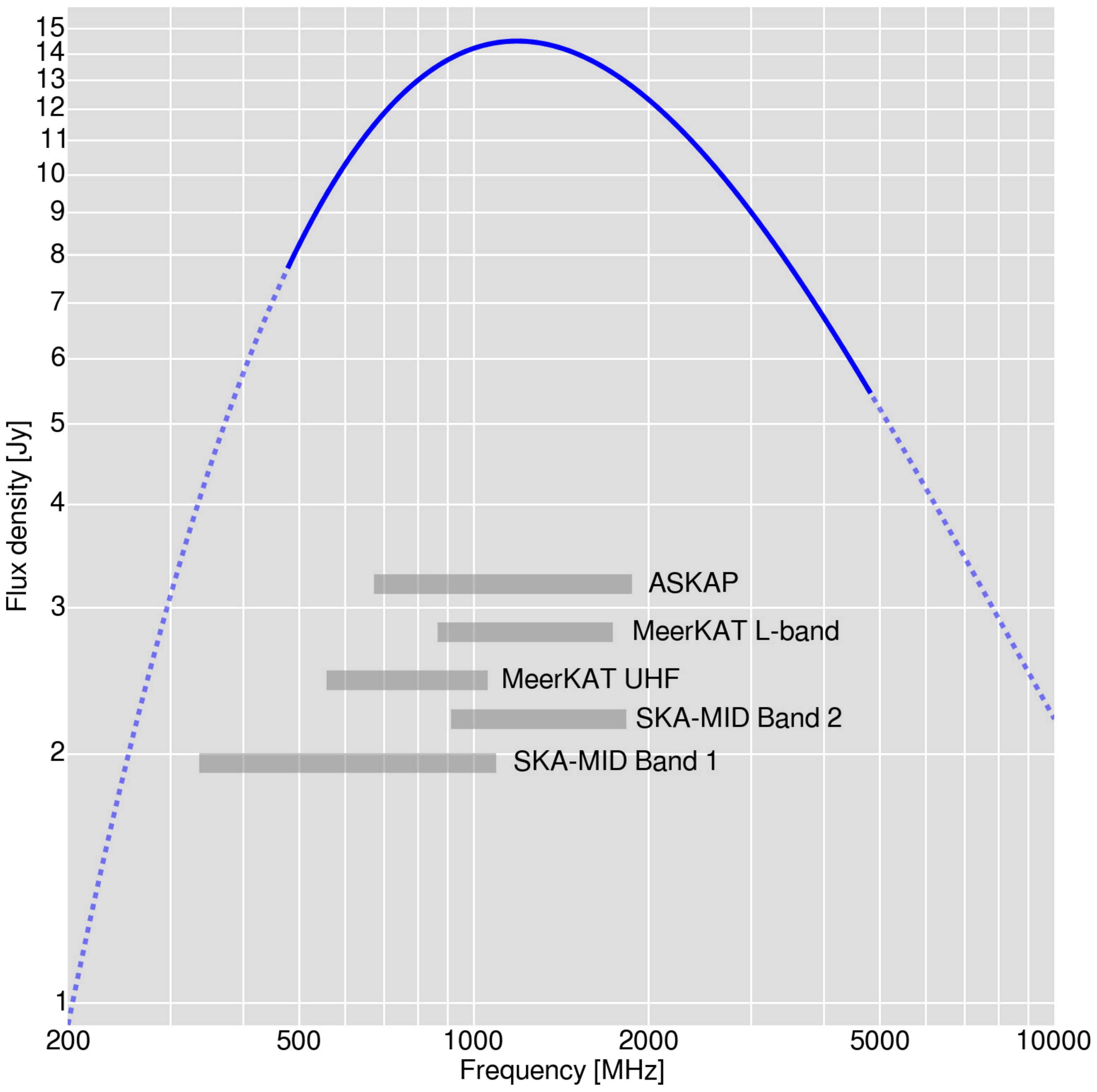}
\caption{Radio spectrum of PKS B1934-638 derived from the model of \citet{reynolds94}. The solid blue curve shows the frequency extent over which measurements were used to derive the polynomial fit to the spectrum, and the dashed line shows the extrapolation of the spectrum using that polynomial. Also marked on the figure are the frequency ranges of ASKAP (0.7 -- 1.9 GHz), MeerKAT's UHF (0.58 -- 1.05 GHz) and L-band ( 0.87 -- 1.67 GHz) receivers, and the assumed ranges of Bands 1 (0.35 -- 1.05 GHz) and 2 (0.95 -- 1.76 GHz) of SKA-MID.}
\label{fig:spectrum}
\end{figure}

Bandpass calibration in the context of radio interferometry is the process of correcting the frequency-dependent instrumental response. The gain of the instrument as a function of frequency is imparted primarily due to the response of electronic components in the signal path. Every independent receiver chain will have a different response, thus each element in the array (and each feed in an individual element) will require a unique bandpass correction, with the general assumption being that the instrumental bandpass is stable in time relative to the duration of a typical observation. Bandpass corrections are generally derived by making a scan of a calibrator source whose spectral and temporal behaviour is well known. A set of antenna-based corrections are derived from the baseline-based visibility measurements by averaging the scan in time and solving for the complex gain in each channel against a model of the calibrator source. In the case of PKS B1934-638 the model is a point-source with a spectrum as shown in Figure \ref{fig:spectrum} \citep{reynolds94}.

Calibration of the bandpass shape is essential to reliably recover the intrinsic structure of astrophysical spectral lines, with experiments demanding high dynamic range requiring a bandpass correction that also achieves an appropriately high signal to noise ratio. Errors in these multiplicative antenna-based bandpass corrections propagate into the baseline spectra of target sources when the calibration is applied. For modern broadband radio interferometers even `continuum' observations require accurate bandpass correction as the broad bandwidths are relied upon for the sensitivity boost that they provide when the whole band is imaged using multi-frequency synthesis techniques \citep[e.g.][]{sault99}.

In the sections that follow we test the suitability of using simple model of PKS B1934-638 for the dish-based mid-frequency component of Phase I of the Square Kilometre Array (SKA-MID), as well as its precursor instruments the Australian Square Kilometre Array Pathfinder \citep[ASKAP;][]{deboer09} and MeerKAT \citep{jonas16}. These instruments have high instantaneous sensitivity and large fields of view due to a combination of technological advances in receiver technology, and the use of dishes in the 12 -- 15 m diameter range. Thus the assumption that a simple point-source calibrator is well-matched to the measurements may not be valid for all observational scenarios due to the contributions to the measurements from other sources in the field. Previous studies have been made on the effects on incomplete sky models during continuum self-calibration \citep[e.g.][]{grobler14}, however here we consider the effect whereby errors are transferred to a target in the spectral domain via calibration tables due to incomplete modelling of a calibrator source. The frequency ranges over which we perform the tests are also marked on Figure \ref{fig:spectrum}. These are 0.7 -- 1.9 GHz \citep[ASKAP;][]{deboer09}, 0.58 - 1.05 GHz and 0.87 -- 1.67 GHz \cite[MeerKAT UHF and L-band respectively;][]{jonas16}, and 0.95 -- 1.76 GHz \citep[SKA-MID Band 2;][]{dewdney13}.

We construct a model of the field sources surrounding PKS B1934-638 using high dynamic range images from ASKAP and the Australia Telescope Compact Array (ATCA). We then use a simulation to test the effect that using an incomplete (central point-source only) model has on the derived bandpass corrections, and how these errors propagate into the target spectral line observations.

\section{Sky model construction}
\label{sec:model}	

We used two observations in order to construct a model of the field around PKS B1934-638. The first of these is an image formed using all of the calibrator scans of PKS B1934-638 taken prior to the Compact Array Broadband Backend \citep{wilson11} upgrade to the ATCA in order to form a deep, narrow-band image of the field at 1.4 GHz. Only arrays in the 6~km configurations were chosen in order to maximise the angular resolution for morphological characterisation of sources close to the primary target. The second data set is a deep ($\sim$12 h) integration of the field using the Boolardy Engineering Test Array \citep[BETA;][]{hotan14} with a frequency range of 711 -- 1015~MHz, for a central frequency of 863 MHz. The image formed from this observation allows the detection of sources out to a radius of approximately 2 deg from the primary target.

\begin{figure}
\centering
\includegraphics[width=\columnwidth]{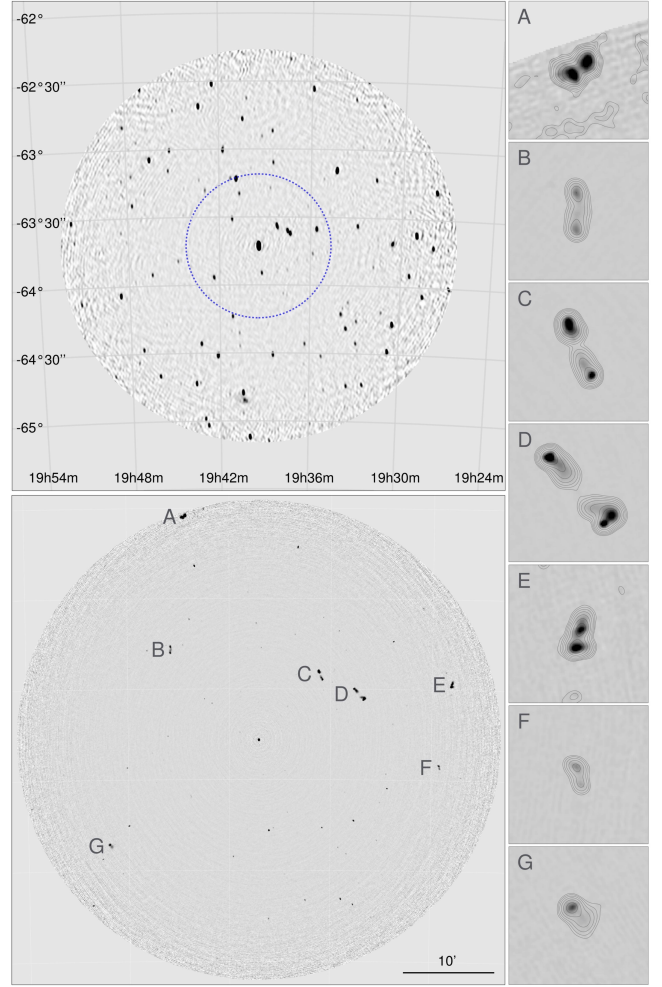}
\caption{The upper-left panel shows the BETA image of the PKS B1934-638 field out to the nominal 10\% radius of the primary beam at 863~MHz. The greyscale is linear and runs from $-$6 mJy~beam$^{-1}$ to 30 mJy~beam$^{-1}$. The dashed circle shows the extent of the ATCA image, cut at the 10\% primary beam level, and shown in the lower left panel. Again the greyscale is linear, running from $-$0.6 mJy~beam$^{-1}$ to 2 mJy~beam$^{-1}$ The seven sources that are not well characterised by single point or Gaussian components are marked A--G, with contour images for each presented in the right hand column. The contour levels are 0.1~$\times$~(3$^{0}$, 3$^{0.5}$, 3$^{1.0}$, 3$^{1.5}$, $\ldots$) mJy~beam$^{-1}$. There are more contours associated with the region around source A as the primary beam correction has raised the image noise towards the edge of the map. Each of the seven sub-images spans 3~$\times$~3 arcminutes.}
\label{fig:maps}
\end{figure}

We made no attempt to modify the standard models used for PKS B1934-638 itself. Self-calibration of the data followed standard procedures, with the BETA data being split into four separate sub-bands. The \citet{reynolds94} model of PKS B1934-638 was used as a starting point for both the BETA and the ATCA data. Self-calibration was performed in several iterations using initial, conservative phase-only solution intervals to prevent the suppression of field sources. Following each iteration, the sky model was expanded to include new components as they emerged in the improved images, until no further improvement to the dynamic range of the images was discernible. Calibration of the ATCA and BETA data was performed using the {\tt DIFMAP} \citep{shepherd97} and {\tt MeqTrees} \citep{noordam10} packages respectively. The final images from which the model catalogue was constructed are shown in Figure \ref{fig:maps}.

A sky model consisting of point and Gaussian components was derived in a hierarchical fashion. Features in the images were decomposed into such components using the PyBDSF source finder \citep{mohan15}, with the exception of the principal component, PKS B1934-638 itself, for which we simply adopted a point source with the polynomial fit to the spectrum derived by \citet{reynolds94}. To this we added the seven sources in the ATCA image found to exhibit complex morphologies, as labeled A--G in Figure \ref{fig:maps}. PyBDSF used between 3 and 16 point or Gaussian components to characterise the extended structures. The BETA observations were used to estimate the total integrated flux densities of each source at 863 MHz and spectral indices were assigned accordingly. The third tier of sources are the point-like features. Positions from the ATCA images were used where available, with total intensity measurements estimated from the BETA image. Attempts to fit spectral shapes between the BETA and ATCA images resulted in implausible spectral indices for several sources, likely due to the resolution mismatch between the observations, so for tier three sources a spectral index ($\alpha$) of $-$0.7 was assumed (i.e.~that of a typical synchrotron radio source), adopting the convention that flux density $S$ is proportional to the observing frequency $\nu$ according to $S~\propto~\nu^{\alpha}$. The sky model at this point consists of 142 point or Gaussian components, not including PKS B1934-638 itself.

\section{Effect of field sources on baseline visibilities}
\label{sec:visibilities}

\subsection{An analytic example}

Consider a point source calibrator at the phase centre with a flux density $S_{0}$, and a second source of flux density $S_{1}$ within the antenna primary beam at coordinate ($x_{1}$,0). The analytic expression for the Fourier transform of these two unresolved sources gives us the visibility function
\begin{equation}
    V(u(t,\nu),v(t,\nu)) = S_{0} + S_{1}e^{-i\phi}
\end{equation}
where
\begin{equation}
    \phi = 2\pi u(t,\nu) x_{1}.
\end{equation}
\noindent
By Euler's formula, the measured intensity is
\begin{equation}
    |V(u(t,\nu),v(t,\nu))|^{2} = S_{0}^{2} + S_{1}^{2} +  2S_{0}S_{1}\mathrm{cos}(-\phi)
\end{equation}
\noindent
and the phase term is:
\begin{equation}
    \mathrm{arg}(V(u(t,\nu),v(t,\nu))) = \mathrm{tan}^{-1}\left(\frac{S_{1}\mathrm{sin}(-\phi)}{S_{0} + S_{1}\mathrm{cos}(-\phi)}\right).
\end{equation}
\noindent
The intensity and phase in this simple example are periodic functions with a time and frequency dependence that is coupled to projected baseline length as well as the distance of the secondary source from the phase centre. For $S_{0} \gg S_{1}$ the phase term will tend to zero, whereas for $S_{0} \rightarrow S_{1}$
\begin{equation}
    \mathrm{arg}(V(u(t,\nu),v(t,\nu))) \rightarrow \mathrm{tan}^{-1}\left(\frac{\mathrm{sin}(-\phi)}{2\mathrm{cos^{2}}(-\phi/2)}\right)
\end{equation}
and the phase term dependence becomes hour-angle dominated. Every off-axis source within the antenna or station primary beam contributes `ripples' to the visibility function in this manner. 

\subsection{Simulation setup}
\label{sec:predict}

Having derived the model in Section \ref{sec:model} we can demonstrate the effect of multiple additional field sources by simulating a realistic set of visibilities, and this process is the cornerstone of the tests that follow. An arbitrary set of visibilities was simulated by generating a {\tt CASA} \citep{mcmullin07} format Measurement Set, containing time and frequency information, and a set of ($u$,$v$,$w$) tracks appropriate for the array being simulated. For the antenna layouts for ASKAP please refer to \citet{gupta08}, and for those of MeerKAT and SKA-MID \citet{heystek15}. Once the Measurement Set was generated the data column was filled with a set of simulated visibilities based on the sky model. Since the sky model derived in Section \ref{sec:model} represents intrinsic brightness measurements the effects of the primary beam must be applied when predicting. This was implemented by applying a simple (cos$^{3}$) voltage beam appropriate for the dish sizes in the array (resulting in a cos$^{6}$ pattern on the sky), which also captures the frequency dependence across the band. Our method includes proper treatment of the differing dish sizes of SKA-MID baselines that involve MeerKAT (13.5 m) and SKA (15 m) dishes. Note that the primary beam effects were not simulated using an image-plane treatment, but by application of a direction- and frequency-dependent Jones matrix during the visibility prediction stage. Since we are only examining the effects of the confusing sources we did not include thermal noise in any simulations. All simulations here and in the sections that follow were performed using the {\tt MeqTrees} software.

\subsection{Effect on visibility measurements}

\begin{figure}
\centering
\includegraphics[width=\columnwidth]{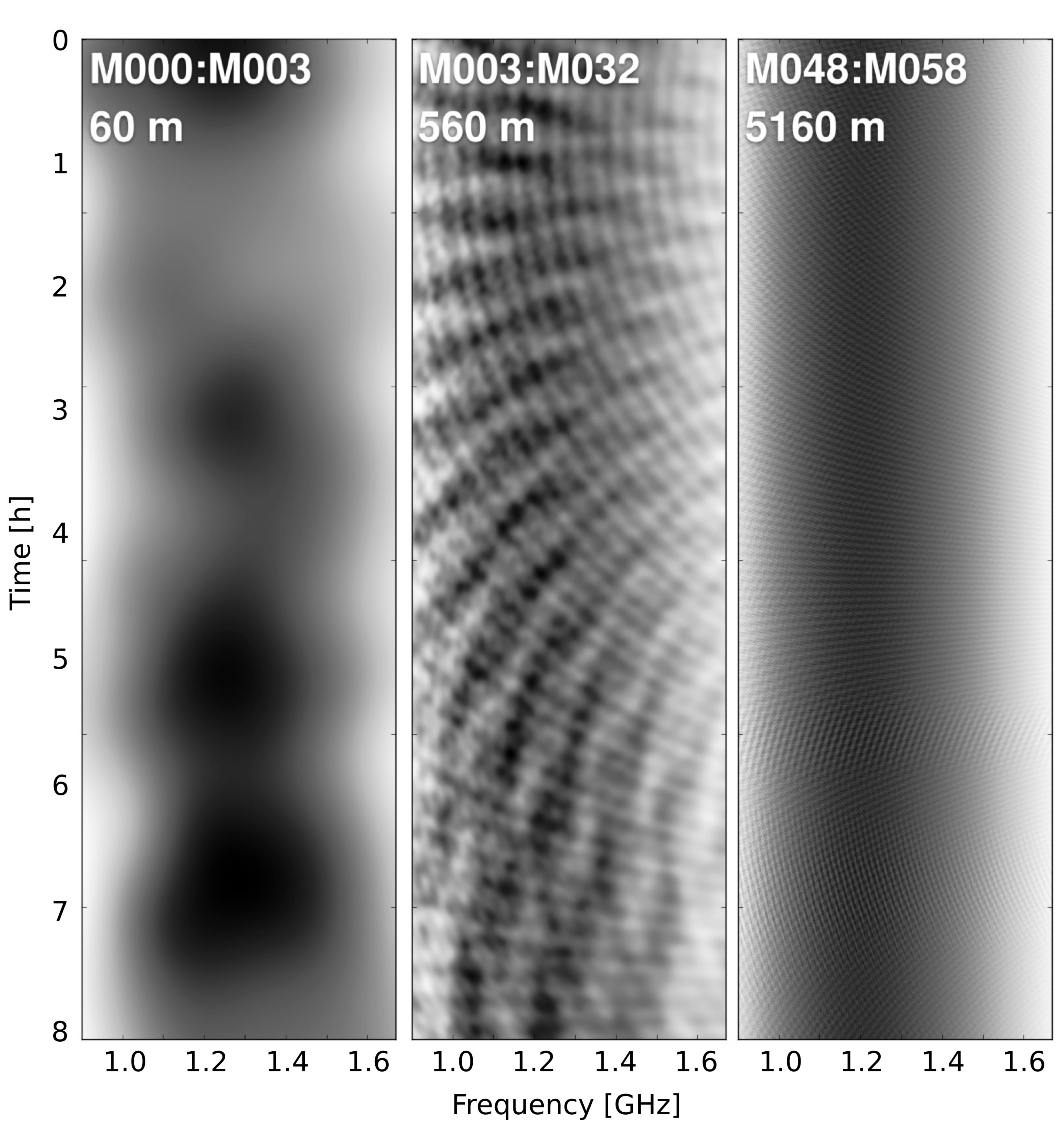}
\caption{Dynamic amplitude spectra for a simulated observation with MeerKAT using the L-band receiver. The vertical axis is time and spans 8 hours, with the horizontal axis being frequency covering 900--1670 MHz. The three panels show three baselines, increasing in factors of approximately ten in length from left to right, with the baseline length and relevant antenna pairs marked on each panel. The underlying point-source model of PKS B1934-638 is modulated by fringe patterns associated with the field sources in a manner that depends on both frequency and baseline length, including the time-dependent projection of the latter. The higher fringe rates associated with longer baselines cause the modulation pattern to be much reduced compared to the shorter baselines. Please refer to the text for further details.}
\label{fig:baselines}
\end{figure}

Figure \ref{fig:baselines} shows the amplitudes of three baselines as a function of time (vertical axis) and frequency (horizontal axis) from an 8-hour MeerKAT track using the full sky model. The simulation covers the full effective frequency range of the L-band receiver (900--1670 MHz). Left to right, the length of the baseline shown increases by a factor of approximately ten between each panel, with the de-projected baseline length indicated together with the relevant antenna pair. The underlying response to PKS B1934-638 is best seen on the right-hand panel. As expected for a point source at the phase centre, the primary calibrator source is stable in time with a spectrum that is curved in frequency, with minimum and maximum amplitudes of 13.49 and 14.49 Jy respectively. The smooth and stable response to PKS B1934-638 is modulated by the presence of the additional sources in the model. The obvious fringe pattern seen in the central panel is primarily introduced by the next brightest pair of sources, the classical doubles labelled C and D with reference to Figure \ref{fig:maps}. The higher fringe rates of the longer baselines cause this pattern to be more effectively smoothed out, however it can still be seen on the 5~km baseline of the right-hand panel. Left to right, the (min, max) values of the amplitudes on each plot are (13.78, 15.54), (13.73, 15.65) and (14.04, 15.35), i.e.~the visibility amplitudes imparted by PKS B1934-638 alone are modulated by up to $\sim$10\% in this case.

The amount of modulation obviously depends on several factors. The projected baseline length causes the fringe pattern of each source to change, and this introduces a time dependence. The frequency dependence exists due to the interplay of two principal effects: (i) intrinsic spectral shapes change the relative contributions of PKS B1934-638 and the field sources to the visibility amplitudes; in particular since PKS B1934-638 turns over at $\sim$1~GHz and the most significant field sources likely have typical synchrotron spectra, the field sources make increasingly significant contributions as the observing frequency is lowered; (ii) the angular size of the primary beam of the antennas increases on the sky with decreasing frequency, further increasing the apparent contribution from off-axis sources.

A typical scan made for bandpass calibration purposes will (depending on the science requirements) last for several minutes. All of the data will be averaged in time, and the measured visibilities will be compared to a predicted set of model visibilities in order to derive antenna-based corrections for each frequency channel. We have briefly demonstrated above the level to which field sources around PKS B1934-638 will affect the visibility function in the case of MeerKAT, with the true response of the instrument (not including thermal noise) during a calibration scan being akin to an appropriate subset of the long track shown, and quite clearly differing from the response to an isolated point source.

\section{Perturbations to the instrumental bandpass corrections}
\label{sec:bandpass}

The plots in this section show the perturbations to the instrumental bandpass corrections that result from assuming a model that consists only of PKS B1934-638. For each of the scenarios under test a set of `observed' visibilities is generated using the process described in Section \ref{sec:visibilities}, featuring the sky model derived in Section \ref{sec:model}. Bandpass corrections are then generated using the {\tt CASA bandpass}\footnote{Version 5.1.1} task, which returns a set of per-channel antenna-based gain corrections that minimise the difference between the `observed' visibilities and the point model. Task defaults were used, with the principal relevant parameters being the minimum of four baselines being present before a solution is attempted for a given antenna, and a minumum signal-to-noise ratio of 3 being required to avoid a solution being flagged. The reference antenna is set to the first one in the Measurement Set, which for all of the instruments considered will be on that is close to the centre of the 
array. The bandpass solutions were not normalised.

\begin{figure*}
\centering
\includegraphics[width=4.7in]{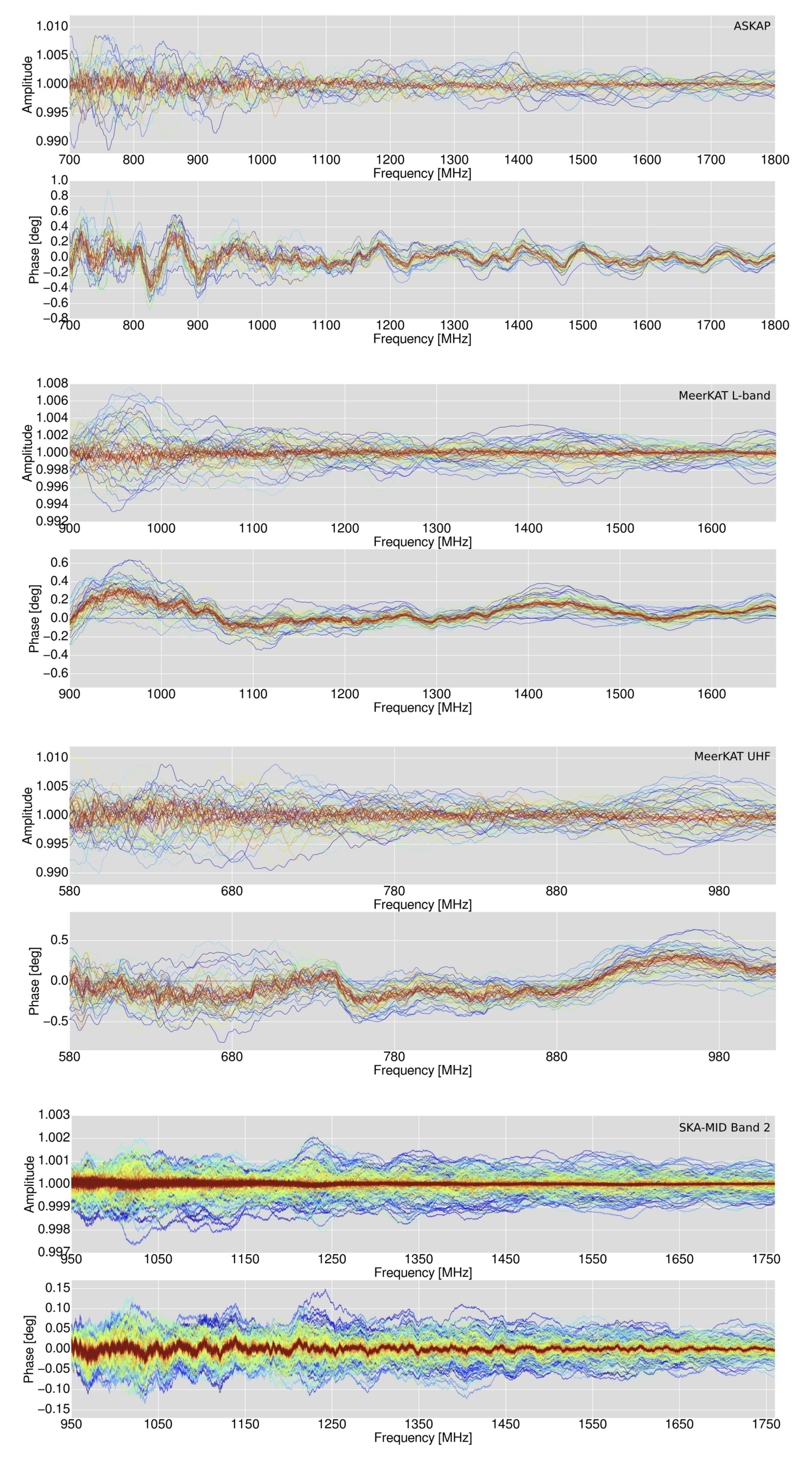}
\caption{Complex (amplitude and phase) bandpass corrections for (top to bottom) ASKAP, MeerKAT L-band, MeerKAT UHF and SKA-MID Band 2 as a function of frequency, derived from 5 minute calibrator scans centred at transit. These plots represent the errors introduced into the bandpass by calibrating with a sky model that consists only of PKS B1934-638. The summed lengths of the (de-projected) baselines that an antenna contributes to have been determined, and are coloured from shortest (blue) to longest (red). Note the differing scales on the y-axes for these panels. Please refer to Section \ref{sec:bandpass} for details.}
\label{fig:bandpass}
\end{figure*}

We generated these simulated corrections for the frequency ranges of the instruments under consideration, namely ASKAP, MeerKAT (L-band and UHF), and SKA-MID (Band 2), and the results are shown in Figure \ref{fig:bandpass}, with the relevant array labeled on each panel. Each of the four scenarios are presented, with a pair of panels separately showing the amplitude and phase of the complex antenna-based bandpass corrections as a function of frequency, for each instrument that we simulate. The colours represent the average deprojected length of the baselines formed with each antenna, running from blue (shortest) to red (longest).

For Figure \ref{fig:bandpass} the simulation is conducted with a bandpass integration time of five minutes, with hour angle coverage that has the target field transiting at the mid point, i.e. the point at which PKS B1934-638 can be observed with the longest projections of baseline length on the sky (given the location of the array on the earth).

Since the true instrumental bandpass in this simulation is a unity response with an amplitude of one and a phase of zero, and given that we do not include the effects of thermal noise, the plots represent the errors imparted to the instrumental bandpass corrections purely due to the effects of using an incomplete sky model for the field.

The main notable features in Figure \ref{fig:bandpass} are intuitively connected to the visibility effects described in Section \ref{sec:visibilities}:
\begin{itemize}
	\item{Errors in the antenna-based bandpass corrections become more severe for antennas that have shorter total baseline lengths. This is due to the higher fringe rates on longer baselines washing out the contribution from confusing sources over the averaging interval used to derive the bandpass corrections.\\}
	\item{The effect becomes more severe with decreasing frequency, where the field sources have an increasing contribution due to the expansion of the primary beam, and the spectral turnover of the primary calibrator source. This is most readily seen by comparing the MeerKAT UHF and L-band plots, where there is both a higher absolute amplitude and phase error, as well as increased spread in the longer baselines for the former.\\}
	\item{Large-scale modulations across the band are likely due to the large number of short spacings in these core-heavy arrays dominating the solutions. SKA-MID has much longer baselines and in greater numbers than either ASKAP or MeerKAT. These serve to mitigate this large scale modulation in the case of SKA-MID. The large number of long baselines will see less contribution from field sources due to fringe rate effects. These two factors serve to reduce both the absolute errors in the bandpass, and flatten the large-scale response in frequency, however the antennas involved in shorter baselines still exhibit the large scale variations.}
\end{itemize}

\begin{figure*}
\centering
\includegraphics[width=4.7in]{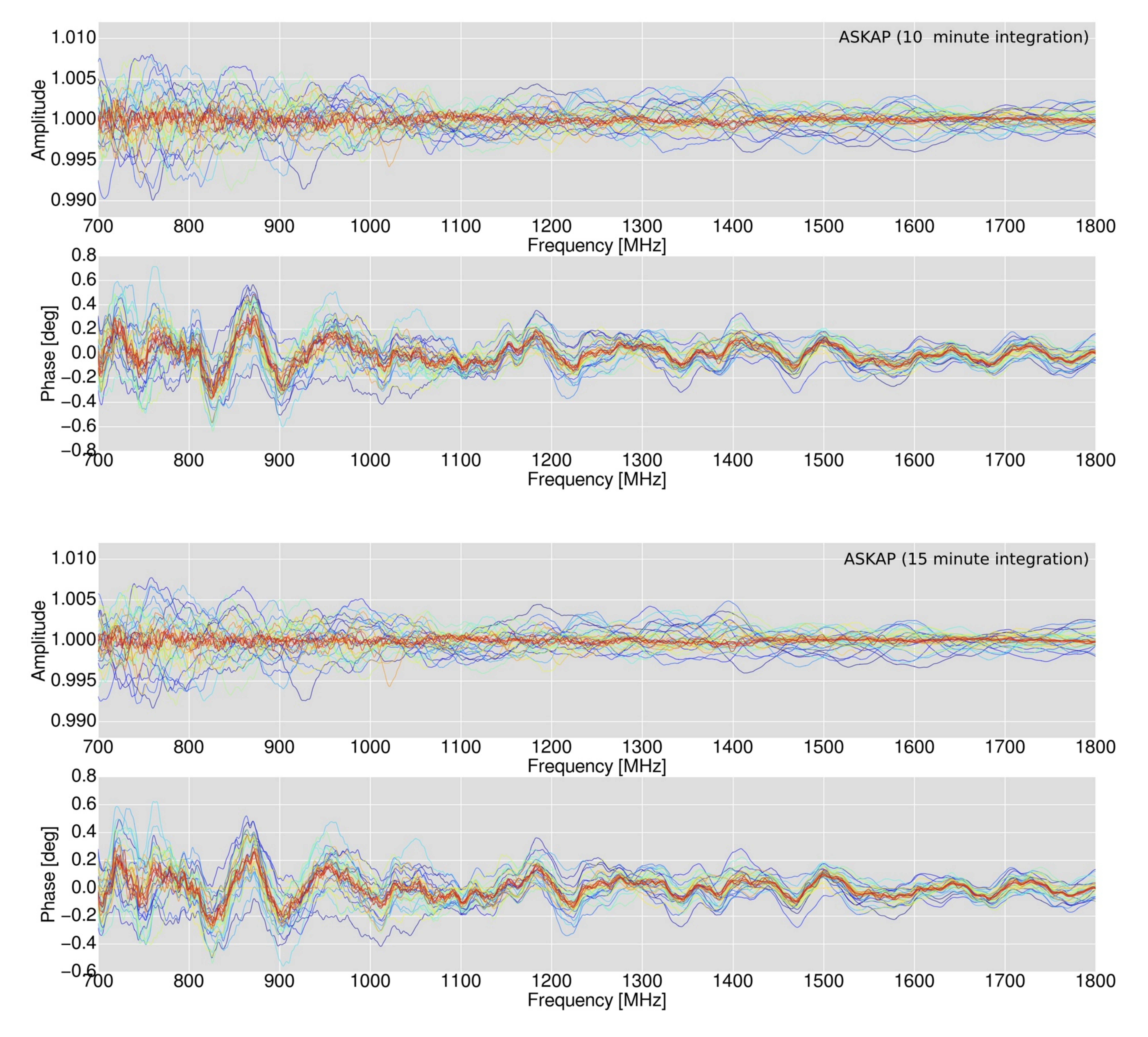}
\caption{Bandpass corrections derived from 10 (upper two panels) and 15 minute (lower two panels) calibrator scans centred at transit for the ASKAP case. These are for comparison to the upper two panels of Figure \ref{fig:bandpass}.}
\label{fig:bandpass_integration}
\end{figure*}

\begin{figure*}
\centering
\includegraphics[width=4.7in]{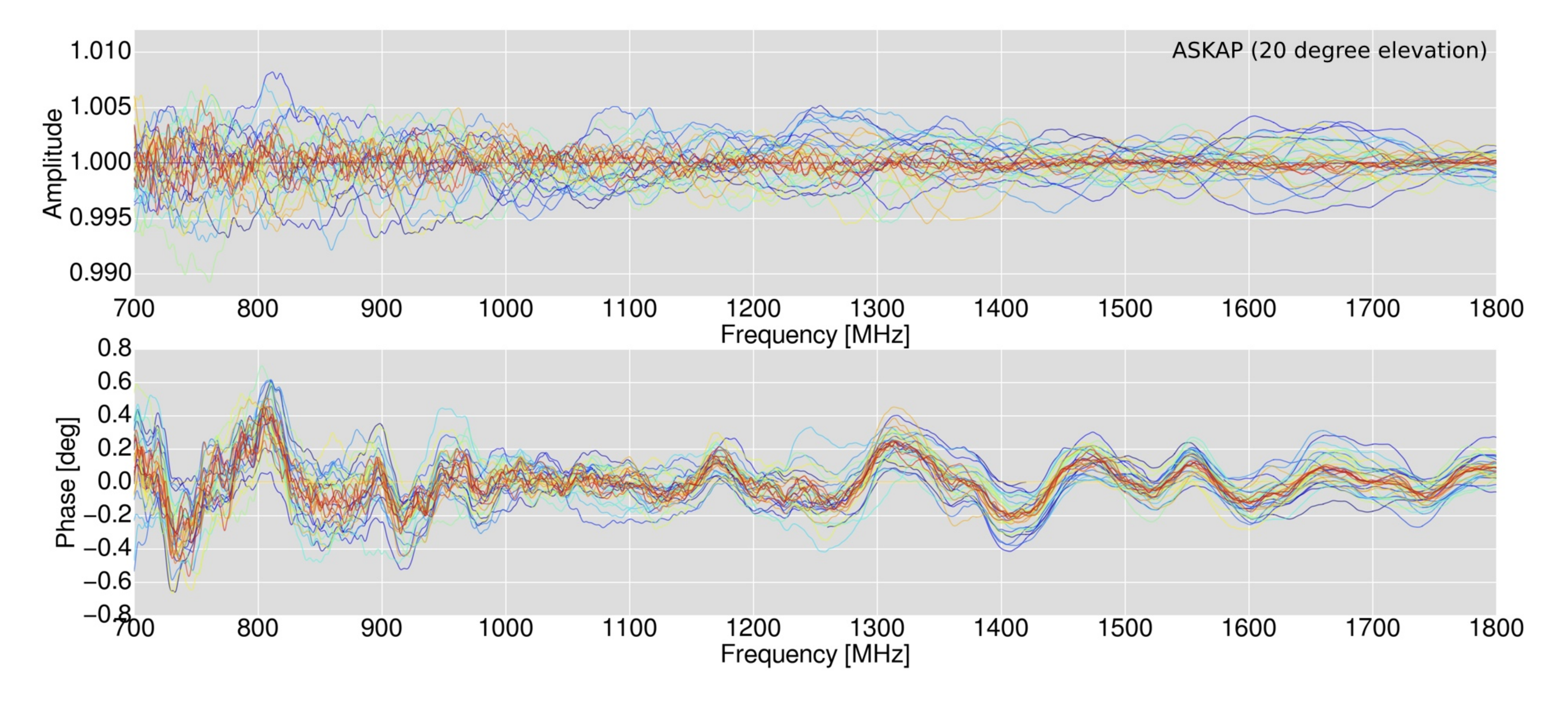}
\caption{Bandpass corrections derived from a 5-minute calibrator scan at 20 deg elevation for the ASKAP case. This is for comparison to the upper two panels of Figure \ref{fig:bandpass}.}
\label{fig:bandpass_elev}
\end{figure*}

Using the ASKAP example, we qualitatively demonstrate the effect of using longer averaging intervals, and different hour angle coverage in Figures \ref{fig:bandpass_integration} and \ref{fig:bandpass_elev}. The longer averaging intervals will affect the bandpass errors due to time-smearing effects reducing the influence of far-field sources, particularly on the long baselines. The hour-angle variations will change the projection of the array on the sky, thereby affecting the effective baseline distribution.

For Figure \ref{fig:bandpass_integration} we repeated the full-band ASKAP simulation presented in Figure \ref{fig:bandpass}, but increased the observation length to 10 (upper amplitude and phase plots) and 15 minutes (the lower pair). Marginal improvement is seen in the absolute errors, and the large scale modulation in frequency also persists. Since the short baselines dominate the solutions (at least in the case of both ASKAP and MeerKAT) this is to be expected. Furthermore, since longer averaging intervals affect not only the visibility amplitudes on the longer baselines but also preferentially wash out sources further from the field centre, the minimal improvement in absolute error from a longer averaging time can be further explained by considering the sky model derived in Section \ref{sec:model}. Referring to Figure \ref{fig:maps}, after PKS B1934-638 the most dominant sources are by far the pair of bright FR-II \citep{fanaroff74} sources, labeled C and D, which are in relatively close proximity (as projected on the sky) to the primary calibrator source. Indeed it is these two sources that contributed mainly to the modulation of the baseline dynamic spectra presented in Figure \ref{fig:baselines}.

We repeated the 5-minute ASKAP simulation from Figure \ref{fig:bandpass}, however this time simulated an observation that occurs when the target field is at an elevation of 20 degrees. The resulting complex bandpass corrections as a function of frequency are shown in Figure \ref{fig:bandpass_elev}, for direct comparison to the corresponding transiting simulation shown in the upper two panels of Figure \ref{fig:bandpass}. Again the improvement in the absolute error is marginal, however the large-scale structure in frequency has changed. This we explain in terms of the different projection of the array onto the sky providing a very different effective baseline distribution compared to the transiting case.

\section{Effect of incomplete model on spectral line observations}
\label{sec:target}

Having shown the level at which an incomplete model introduces errors into the bandpass correction, we now demonstrate how these errors propagate into the target visibilities when the bandpass is applied, and thus the errors introduced into the spectrum of a target source. To show this we generated model visibilities consistent with a 1 Jy flat spectrum point source at the phase-centre (zero phase response for all baselines), and applied the appropriate corrupted bandpass table using the {\tt CASA applycal} task. We then took the perturbed (`corrected') visibilities and examined their vector-averaged spectrum. The results are shown for ASKAP, MeerKAT UHF, and MeerKAT L-band in Figure \ref{fig:target1}, and for SKA-MID Band 2 in Figure \ref{fig:target2}, expressed as a percentage error in the true source spectrum.

\subsection{Effects of hour-angle coverage and duration of calibrator scan}
\label{sec:hourangle}

\begin{figure}
\centering
\includegraphics[width=\columnwidth]{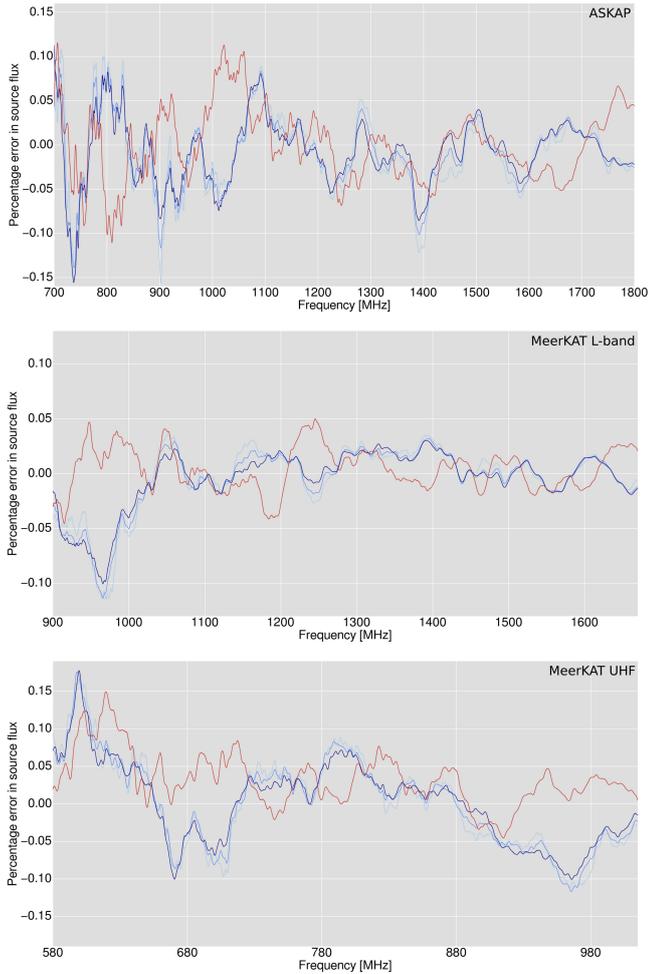}
\caption{Corruptions to the vector-averaged visibility spectrum of a flat-spectrum point source at the phase centre due to bandpass calibration from the incomplete model. The spectra are expressed as the percentage error of the target flux density. These are shown top to bottom for ASKAP, MeerKAT L-band, and MeerKAT UHF band. The blue traces (light to dark) are for bandpass calibration scans of 5, 10 and 15 minutes at zenith. The single red trace is for a bandpass calibration scan of 5 minutes duration at an elevation of 20 degrees.}
\label{fig:target1}
\end{figure}

The effects of hour-angle (elevation) and averaging interval during the bandpass observation are captured in Figure \ref{fig:target1}. The blue and red traces represent transit and 20 degree elevation scenarios respectively, and for the transit scenario the different shades of blue represent integration times of 5, 10 and 15 minutes (light to dark). The integration time effect for the 20 degree case is omitted for clarity. 

As expected, the behaviour of these plots mimics those of the bandpass tables, with the visibility per baseline corrupted by the multiplicative bandpass amplitude errors from the two elements involved on a per-channel basis. Broad features (widths of $\sim$20 MHz) are artificially introduced into the spectra, corrupting the large-scale behaviour of the spectral baseline response at the $\sim$0.1\% level for ASKAP and MeerKAT, dropping to the $\sim$0.01\% level for SKA-MID. Comparing the red curves to the blue curves shows that the morphology of this large-scale corruption is coupled to the array layout (or projected array layout as modified by elevation effects). Increasing the length of the calibrator scan primarily affects the higher-frequency structure in the spectral errors, mainly due to the longer averaging interval suppressing the contribution of off-axis sources to the longer baselines.

The large-scale error in the source spectrum due to incomplete modelling of the bandpass calibrator field is a direction-independent effect that will apply equally to all sources in an image cube, and would be difficult to remove using further continuum subtraction techniques as in all cases it would not be well-described by a low order polynomial. 

\begin{figure}
\centering
\includegraphics[width=\columnwidth]{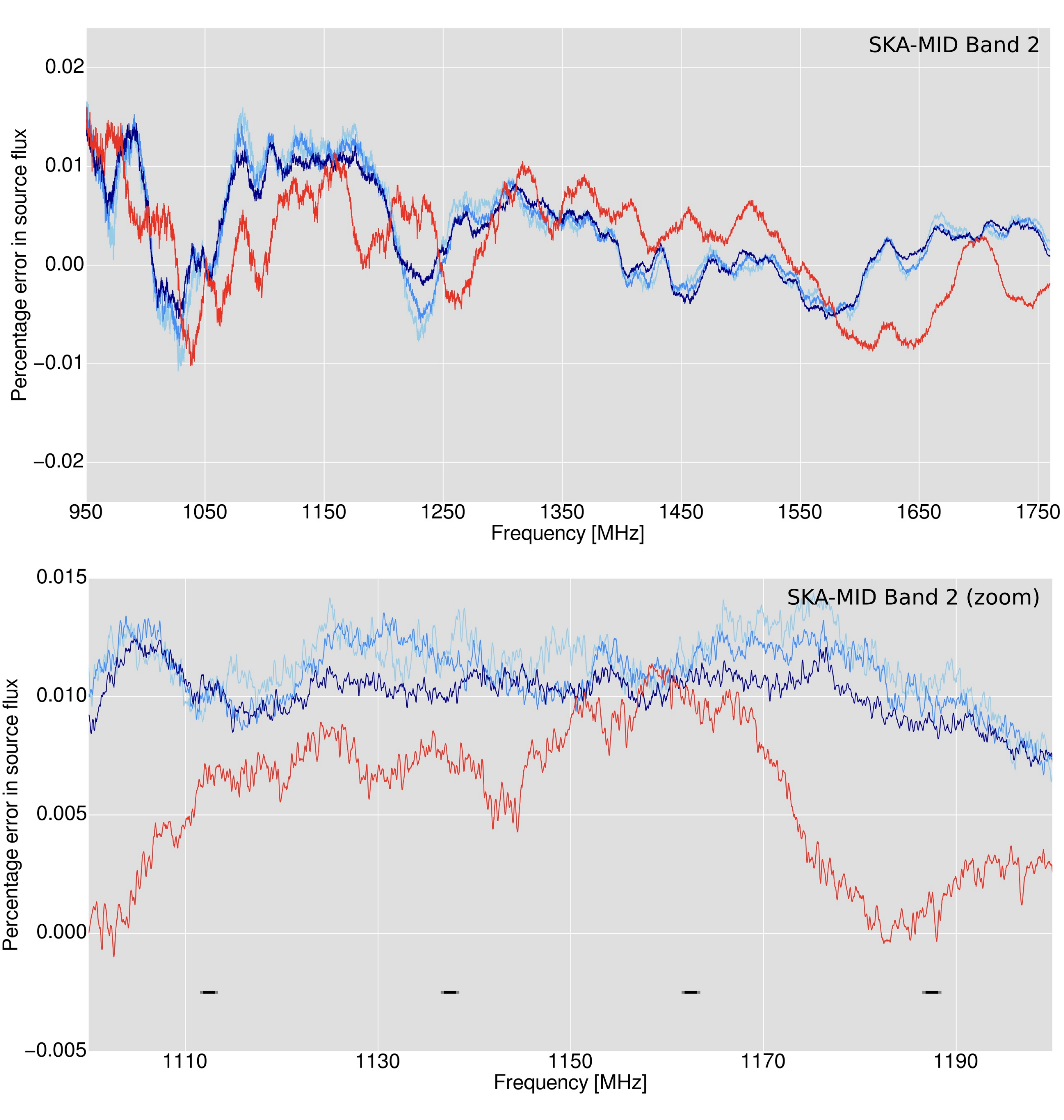}
\caption{These panels are as per Figure \ref{fig:target1}, but for Band 2 of SKA-MID, with the lower panel showing a zoomed region of the upper panel to demonstrate the high-frequency perturbations to the spectrum. The dashed lines on the lower panel show the mean and standard deviation of the line widths for HI discs at that redshift, as predicted by \citet{obreschkow09}.}
\label{fig:target2}
\end{figure}

The higher frequency structure in the introduced errors is perhaps more insidious. Although the fractional error is smaller for this component, the widths are narrower and could mimic genuine spectral line emission or absorption features in high dynamic range line observations. This is particularly true for the SKA-MID case, as shown in Figure \ref{fig:target2}. The lower panel shows a 100 MHz region of the full spectrum shown in the upper panel. The dark lines show the mean line widths for HI discs at the corresponding redshift for four frequencies in the zoom region, with the grey extensions showing $\pm$1 standard deviation, as drawn from the extragalactic sky simulation of \citet{obreschkow09}. Interpretation of spectra in the presence of these high frequency corruptions should be made with caution, in e.g.~SKA-MID line experiments with dynamic range demands of $\sim$10$^{5}$ and above. This is an atypically high requirement for spectral line experiments, however could be a potential contaminant of HI lines peaking at a few tens of mJy in for example the Ultra Deep SKA-MID survey outlined by \citet{blyth15}.

An important real-world consideration that we have hiterto ignored is the tricky issue of continuum subtraction. The spectral corruptions we investigate here will also be present in the (generally smooth) spectra of the radio continuum emission, and thus in addition to apparent spectral line corruptions, continuum subtraction under the assumption of a smooth model also becomes inappropriate. As an example, the continuum emission of a typical star forming galaxy with a HI-to-continuum flux density ratio of 2:1 will exhibit corruptions at a comparable level to the examples above.

\subsection{Effects of imaging weights}

\begin{figure*}
\centering
\includegraphics[width=7in]{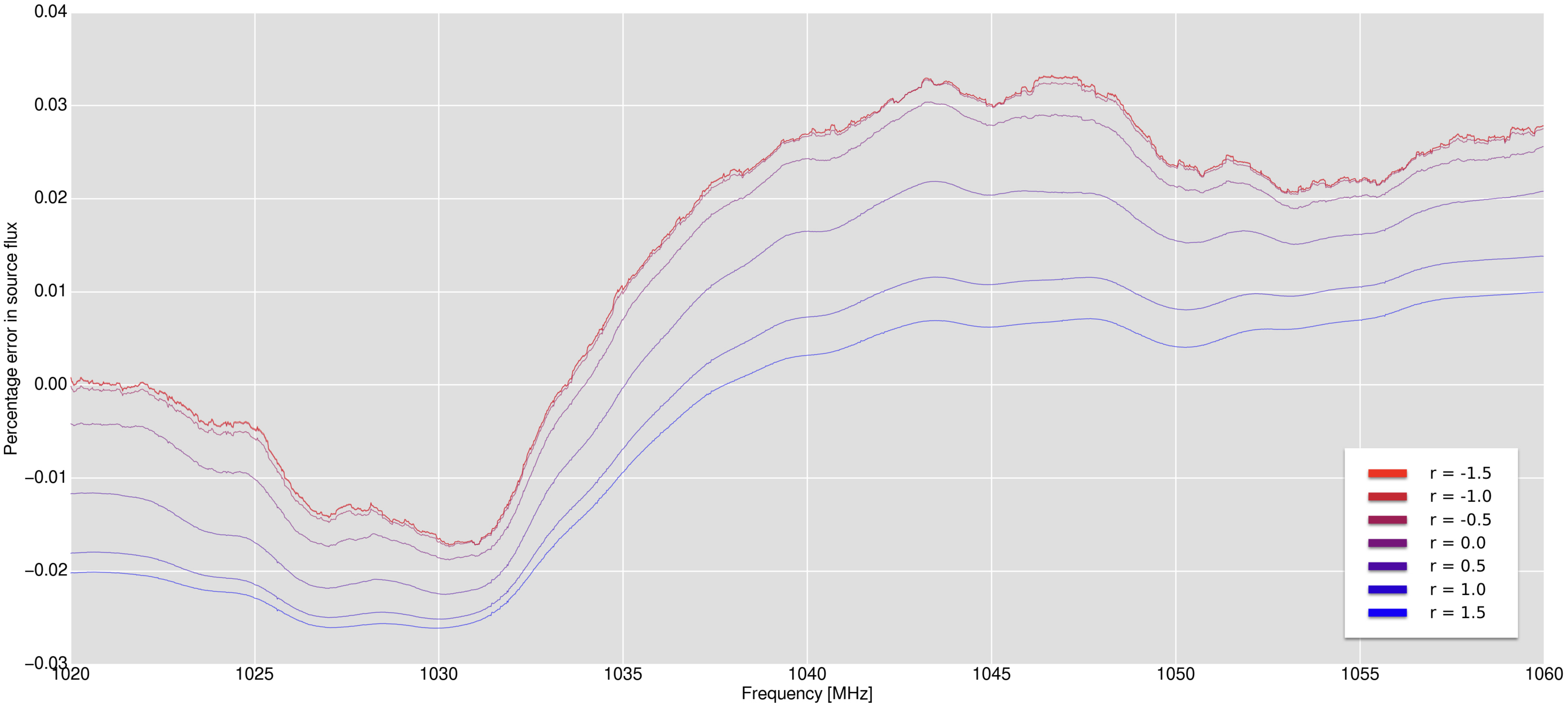}
\caption{A subset of the MeerKAT L-band simulation showing the percentage error in the spectrum of a target source as a function of frequency, for seven different values of the robust parameter. The robust parameters are colour coded as per the legend.}
\label{fig:robusts}
\end{figure*}

The core-dominated antenna layouts for the three arrays considered in our simulations mean that for most imaging applications a significant downweighting of the core spacings must be applied in order to condition the point spread function (PSF) for reliable deconvolution. This is generally achieved by tuning of the robustness parameter ($r$) \citet{briggs95}. The vector-averaged visibility spectra derived above are equivalent to imaging with purely natural weighting of the visibilities, so here we investigate how the incomplete sky model affects target spectra for a range of different weighting schemes.

The robustness parameter facilitates a continuous transition between natural weighting ($r$~=~2.0), for which every visibility has unity weighting, and uniform weighting ($r$~=~-2.0) for which every grid cell in the ($u$,$v$) plane has unity weighting. Thus for $r$~$<$~2.0 the inevitably higher density of samples close to the origin of the ($u$,$v$) plane means that shorter spacings are preferentially downweighted. The principal motivation is to trade off sensitivity against angular resolution or PSF sidelobe levels. Since we have established that use of an incomplete calibrator model imparts errors that are directly coupled to the intrinsic (or projected) baseline distributions of the arrays, we can expect the robustness parameter to directly modify the behaviour of the spectral perturbations.

The process for testing this is as above, i.e. a corrupted bandpass table is generated and applied to a simulated set of target visibilities. At the final stage, instead of extracting the corrupted spectrum directly from the visibilities, an image cube is produced using {\tt CASA}'s {\tt clean} task, and the spectrum is measured from the cube. The sky in our simulations consists only of a point source at the phase centre, thus wide-field imaging corrections are not required. However in order to capture the effects of visibility weighting in a realistic manner\footnote{For most imaging software the resolution (and PSF sidelobe levels) for any weighting scheme other than natural will depend on the image size (in pixels) and the angular extent of each pixel, as these parameters subsequently set the size and resolution of the grids in the ($u$,$v$) domain. Weighting schemes other than natural use cell-based rather than visibility-based weights, and so adjusting the ($u$,$v$) grid properties changes how the visibilities are distributed amongst the cells.}, we use an image size of 8,192~$\times$~8,192 pixels, each of which is 1.5$''$~$\times$~1.5$''$. Wide-band deconvolution schemes are also not required, as we are treating each channel independently in order to obtain a full spectral measurement, and indeed given the properties of the simulated sky, the spectrum is measured directly from the brightest pixel in the dirty (i.e. not deconvolved) cube.

The imaging process is repeated for $r$ values of -1.5, -1.0, -0.5, 0.0, 0.5, 1.0 and 1.5. Since this is a somewhat computationally expensive simulation we test only the case of MeerKAT L-band, and then only for 2,000 frequency channels between 1020 and 1060 MHz. Again, the extracted spectrum is recast in terms of the percentage error from the true source spectrum, and the results for the seven different weighting schemes are shown in Figure \ref{fig:robusts}. The different values are colour-coded as indicated in the legend. As $r$ values tend towards uniform weighting the high-frequency ripples in the perturbations become more pronounced, as expected for the effect being coupled to the longest baselines in the array. This also suggests that the magnitude of the high frequency ripples highlighted for SKA-MID in Section \ref{sec:hourangle} represent the best case scenario, as for just about all practical imaging applications the core-spacings will have to be suppressed, significantly so for SKA-MID science cases that require sub-arcsecond angular resolution. 

The overall characteristics of the low-frequency component of the introduced spectral errors do not significantly change with varying $r$ parameters. The maximum-minimum error in the examples shown here changed by a few tens of percent, however this is probably irrelevant given the more significant effect that changing the hour-angle / elevation of the calibrator scan has on this particular aspect of the spectral perturbations.

\section{A real-world test case}
\label{sec:realworld}

\begin{figure*}
\centering
\includegraphics[width=7in]{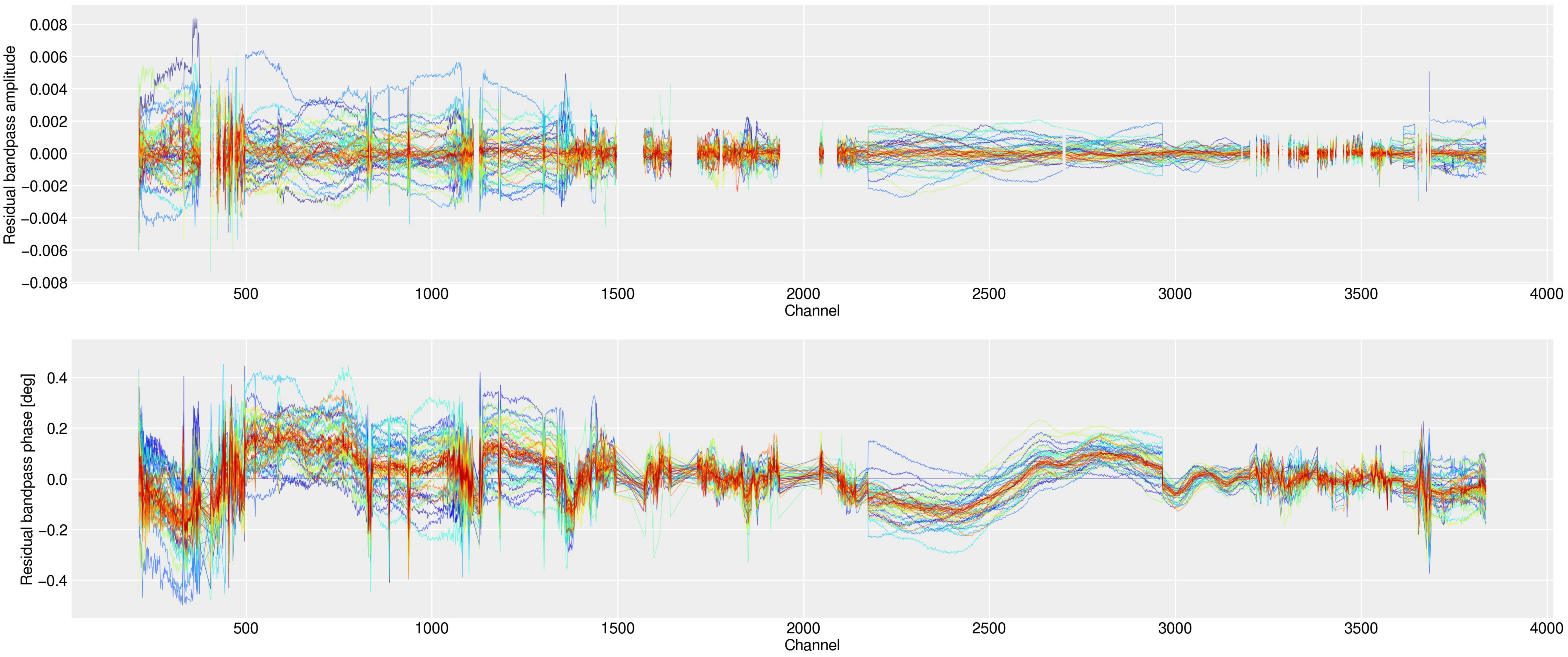}
\caption{Residual bandpass corrections (amplitude and phase) for a MeerKAT L-band observation of PKS B1934-638, showing the difference in the solutions obtained when using a standard sky model consisting only of the central point source versus calibration performed using the more complete sky model we derived in Section \ref{sec:model}. The residual bandpass structures are in remarkably good agreement with those seen in the MeerKAT L-band panels of Figure \ref{fig:bandpass}. This validates both the simulation framework as well as its results. Colour coding on this plot is as per Figure \ref{fig:bandpass}. Please refer to Section \ref{sec:realworld} for further details.}
\label{fig:realbandpass}
\end{figure*}

Our simulation framework and the results derived in this paper are validated using a calibration scan of PKS B1934-638 from the MeerKAT telescope. The observation used 55 antennas, with an on-source time of 8 minutes, observed at an elevation angle of approximately 38 degrees. The correlator was configured to deliver 4,096 frequency channels. Flagging was done using the {\tt CASA flagdata} task. Autocorrelations and visibilities with amplitudes of exactly zero were discarded. The low-gain edges of the band (850--900 MHz and 1658--1800MHz) were flagged. The initial integration (8 seconds) was discarded to mitigate against incomplete slewing and antenna settling time, and visibility amplitudes greater than 100 were flagged. Regions with strong persistent radio frequency interference (RFI; 944--947 MHz, 1160--1310 MHz and 1476--1611 MHz, primarily due to geolocation satellites) were flagged for baselines shorter than 1000 metres. The {\tt rflag} and {\tt tfcrop} algorithms within {\tt flagdata} were then used to automatically identify and remove residual RFI from the remaining visibilities using the default settings. Following this a bandpass correction was derived and applied using a model containing only PKS B1934-638. The two auto-flagging algorithms were then re-run on the residual (corrected - model) visibilities, and the initial bandpass corrections were discarded. The bandpass corrections were then repeated by calibrating against a model that consisted only of PKS B1934-638 and then a set of model visibilities generated in accordance with the simulation method (Section \ref{sec:predict}). The differences between the two resulting bandpass tables are shown in Figure \ref{fig:realbandpass}. The level and structure of the residual bandpass differences are in excellent agreement with those predicted by the simulation, and shown in the MeerKAT L-band case in Figure \ref{fig:bandpass}. Some discontinuities are present which are likely due to residual RFI. 

\section{Conclusions}
\label{sec:discussion}

The general assumption when calibrating radio interferometer data is that calibrator sources are isolated point sources at the phase centre. However the instantaneous sensitivity and wide fields of view of modern radio telescopes may make this assumption inappropriate for many observational scenarios. We have investigated how unmodelled field sources surrounding the typical southern-sky primary calibrator PKS B1934-638 introduce errors into the bandpass corrections, and subsequently the spectra of science targets for observations with ASKAP, MeerKAT and SKA-MID. 

The bandpass tables derived from an incomplete model of the PKS B1934-638 field when transferred to the visibilities of the science target impart spectral corruptions with a range of scales. These corruptions are multiplicative at the antenna level, and thus affect the spectra of all sources in the field (including broad-band continuum emission) at a level that is commensurate with their brightness.

The behaviour of the errors is coupled to the array layout, with the shorter baselines of the array governing the behaviour of the lower frequency corruptions, and the longer baselines giving rise to higher frequency components. The projection of the baselines onto the sky also means that the error patterns have a strong elevation dependence. Increasing the integration time of the calibrator scan (within reasonable limits) preferentially reduces the higher frequency components, as the higher fringe rates of longer baselines effectively suppresses off-axis sources via smearing effects. The use of weighting schemes that tend towards uniform however amplifies these high frequency errors, as more weight is given to the longer baselines in this regime.

With reference to Figures \ref{fig:target1} and \ref{fig:target2} the broad, low frequency component of the spectral corruptions would need to be mitigated to achieve spectral baseline accuracies of greater than approximately 1 part in 1,000 for the precursor instruments, and 1 part in 10,000 for SKA-MID. Continuum subtraction using the typical method of a low-order polynomial fit along the frequency axis would likely be insufficient. The higher frequency component is particularly evident in the SKA-MID simulation. Although the high frequency perturbations exist on scales comparable to the line widths of HI discs at the cosmological redshifts the SKA and its precursors aim to probe, the amplitude of this component is small enough such that is only likely to affect the brighter HI lines in the ultra deep tiers of SKA-MID surveys. 

To provide some brief realistic examples (with $\eta$ being an efficiency term, and $T_{sys}$ being the assumed system temperature):
\begin{itemize}
	\item{{\bf A hypothetical local HI observation with MeerKAT:} A 20~h on-source integration at 1.4~GHz, with 22 kHz channels and $T_{sys}/\eta$~=~22~K has a naturally-weighted channel noise of 0.11 mJy beam$^{-1}$. Thus the 0.1\% maximum bandpass error equals the image noise for a 110 mJy continuum source.\\}
	\item{{\bf A hypothetical deep HI observation at z$\simeq$0.3 with MeerKAT:} A 200~h on-source observation at 1.1~GHz, with $T_{sys}/\eta$~=~23~K and 100~kHz channels has a naturally-weighted channel noise of 18 $\mu$Jy beam$^{-1}$. In this example the 0.1\% maximum bandpass error equals the image noise for a 18 mJy continuum source, something that is relatively common given the field of view.}
\end{itemize}
	
The straightforward way to mitigate all of the issues we have investigated here is of course to expand the models of this field, and indeed all standard calibrator fields. This recommendation is supported by the results of Section \ref{sec:realworld}, which shows the difference between a bandpass table generated with a single point source model and one generated with our full-field model in the case of a real MeerKAT L-band calibration scan of PKS B1934-638. The residual structure is remarkably similar to the predicted corruptions for this scenario shown in Figure \ref{fig:bandpass}. 

Initial results from MeerKAT's UHF system (Hugo, priv. comm.) suggest that the contribution from very strong sources in the sidelobes of the primary beam is significant, and this is an effect that our simulations do not capture due to the simple primary beam model. The use of full-field models that include strong sources in the sidelobes of the antenna primary beam patterns will thus be particularly important for MeerKAT UHF and SKA-MID Band 1, the latter of which we have not considered here. An apparent sky model constructed via an observing campaign for each precursor instrument and the SKA itself would detach any uncertainties in the primary beam model from the problem, insofar as the primary beams of alt/az mounted dishes exhibit azimuthal symmetry. Even simply including sources C and D (Figure \ref{fig:maps}) will greatly reduce the issue at L-band / SKA Band 2 frequencies, as these two sources dominate the visibility perturbations for all scenarios considered here.

Another approach we have not investigated here is the use of ($u$,$v$) range selection to remove the more error-prone shorter baselines from the calibration procedure. This has been shown to be effective during self-calibration of arrays with high concentrations of antennas in the core, where the shorter baselines may pick up large-scale emission features that are difficult to model via deconvolution-based techniques.

Finally, we note that we have only examined this effect in the context of the PKS B1934-638 field. A cursory examination of other fields has shown that the errors resulting from an incomplete model of this field are comparatively small. For example, typical MeerKAT calibration observations of PKS 0407-65 and 3C 286 close to transit will introduce errors of up to 1\% (a factor of a few tens larger) due to incomplete modelling. The conclusion drawn in the previous paragraph naturally applies to all primary calibrator fields. 

\section*{Acknowledgements}
\addcontentsline{toc}{section}{Acknowledgements}

We thank the anonymous referee and the MNRAS editorial staff for their feedback which has significantly improved this manuscript. The Australia Telescope Compact Array is part of the Australia Telescope National Facility which is funded by the Australian Government for operation as a National Facility managed by CSIRO. The Australian SKA Pathfinder is part of the Australia Telescope National Facility which is managed by CSIRO. Operation of ASKAP is funded by the Australian Government with support from the National Collaborative Research Infrastructure Strategy. ASKAP uses the resources of the Pawsey Supercomputing Centre. Establishment of ASKAP, the Murchison Radio-astronomy Observatory and the Pawsey Supercomputing Centre are initiatives of the Australian Government, with support from the Government of Western Australia and the Science and Industry Endowment Fund. We acknowledge the Wajarri Yamatji people as the traditional owners of the Observatory site. The MeerKAT telescope is operated by the South African Radio Astronomy
Observatory, which is a facility of the National Research Foundation, an agency of the Department
of Science and Innovation. IH acknowledges support of the Oxford Hintze Centre for Astrophysical Surveys which is funded through generous support from the Hintze Family Charitable Foundation. This project has received funding from the European Research Council (ERC) under the European Union's Horizon 2020 research and innovation programme (grant agreement no. 679627; project name FORNAX). This research has made use of NASA's Astrophysics
Data System. Some figures in this paper were created using the Python package APLpy, an open-source plotting package for Python \citep{robitaille12}.











\bsp	
\label{lastpage}
\end{document}